\documentclass[12pt,a4paper]{article}
\usepackage[british]{babel}
\usepackage{epsfig}
\begin{document}
\date{}

\title{
{\vspace{-20mm} \normalsize
\hfill \parbox[t]{50mm}{DESY 05-085 }}\\[4em]
 Updating algorithms with multi-step \\
 stochastic correction               \\[1.5em]}
\author{I.\ Montvay and E.\ Scholz \\[1em]
        Deutsches Elektronen-Synchrotron DESY  \\
        Notkestr.\,85, D-22603 Hamburg, Germany}

%%%%%%%%%%%%%%%%%%%%%%%%%%%%%%%%%%%%%%%%%%%%%%%%%%%%%%%%%%%%%%%%%%%%%%%%        
\newcommand{\be}{\begin{equation}}                                              
\newcommand{\ee}{\end{equation}}                                                
\newcommand{\half}{\frac{1}{2}}                                                 
\newcommand{\rar}{\rightarrow}                                                  
\newcommand{\lar}{\leftarrow}
\newcommand{\LCB}{\raisebox{-0.3ex}{\mbox{\LARGE$\left\{\right.$}}}
\newcommand{\RCB}{\raisebox{-0.3ex}{\mbox{\LARGE$\left.\right\}$}}}
\newcommand{\U}{\mathrm{U}}
\newcommand{\SU}{\mathrm{SU}}
\newcommand{\bteq}[1]{\boldmath$#1$\unboldmath}
%%%%%%%%%%%%%%%%%%%%%%%%%%%%%%%%%%%%%%%%%%%%%%%%%%%%%%%%%%%%%%%%%%%%%%%%

\maketitle

\abstract{\normalsize
 Nested multi-step stochastic correction offers a possibility to
 improve updating algorithms for numerical simulations of lattice
 gauge theories with fermions.
 The corresponding generalisations of the two-step multi-boson (TSMB)
 algorithm as well as some applications with hybrid Monte Carlo (HMC)
 algorithms are considered.}

%%%%%%%%%%%%%%%%%%%%%%%%%%%%%%%%%%%%%%%%%%%%%%%%%%%%%%%%%%%%%%%%%%%%%%%%
\section{Introduction}\label{sec1}
 The main task in numerical Monte Carlo simulations of lattice gauge
 theories with fermions is to evaluate the (ratio of) fermion
 determinants appearing in the Boltzmann weight for the gauge fields.
 The idea of the stochastic (``noisy'') correction \cite{KENKUT} is
 to prepare a new proposal of the gauge configuration during updating
 by some approximation of the determinant ratio and accept or reject
 the change based on a stochastic estimator.
 This ``stochastic correction step'' takes care of the deviation of
 the approximate determinant ratio from the exact one.

 In multi-boson updating algorithms \cite{LUSCHER} it is natural to
 introduce a stochastic correction step in order to correct for the
 deviations of the applied polynomial approximations.
 In special cases it is possible to perform the correction by
 an iterative inverter \cite{BORFOR}.
 More generally, the correction step can be based on successively
 better polynomial approximations, as in the two-step multi-boson
 (TSMB) algorithm \cite{TSMB}.
 A suitable way to obtain the necessary polynomial approximations is to
 use a recursive scheme providing least-square optimisation
 \cite{POLYNOM1,POLYNOM2}.
 Based on this stochastic correction scheme, the TSMB updating algorithm
 has been successfully applied in several numerical simulation projects
 both in supersymmetric Yang Mills theory (see \cite{SYMREV} and
 references therein) and in QCD (see, for instance,
 \cite{DENSE,NF2TEST,TWIST,DBW}).

 In the present paper we generalise the idea of stochastic correction
 into a scheme of nested successive corrections based on polynomial
 approximations with successively increasing precision.
 (A similar ``multi-level Metropolis'' scheme has been proposed in
 Ref.~\cite{HASENBUSCH,MULTIMET}.)
 In the next section we consider {\em multi-step multi-boson}
 algorithms.
 The last section is devoted to different possibilities for combining
 multi-step stochastic correction with variants of the hybrid Monte
 Carlo (HMC) updating algorithm \cite{HMC}.
 In particular, optimised HMC algorithms based on mass preconditioning
 \cite{HASENBUSCH,HASENBUSCH-JANSEN} and polynomial hybrid Monte Carlo
 (PHMC) algorithms \cite{PHMC} are considered.

%%%%%%%%%%%%%%%%%%%%%%%%%%%%%%%%%%%%%%%%%%%%%%%%%%%%%%%%%%%%%%%%%%%%%%%%
\section{Multi-step multi-boson algorithms}\label{sec2}

 The multi-step multi-boson (MSMB) algorithm is a generalisation of the
 TSMB updating algorithm.
 Therefore, let us briefly recapitulate the basics of TSMB.
 Let us assume that the determinant of the Hermitian fermion matrix
 $Q=Q^\dagger$ is positive, at least on most of the gauge configurations
 occurring with non-negligible weight in the path integral.
 In this case the sign of the determinant can either be neglected or
 taken into account by reweighting on an ensemble of configurations
 obtained by updating without the sign.
 (If the sign of the determinant plays an important role then there is
 a ``sign problem'' which cannot be dealt with by a straightforward
 Monte Carlo simulation procedure.)
 Without the sign the determinant factor in the Boltzmann weight of
 the gauge configurations is
\be\label{eq01}
|\det Q|^{2\alpha} = \left( \det Q^2 \right)^\alpha \ , 
\ee
 where in case of $N_f$ mass-degenerate Dirac-fermion flavours we have
 $\alpha=\half N_f$.
 (Note that for a Majorana fermion $\alpha=\frac{1}{4}$.)
 Of course, for several fermion flavours with different masses there 
 are several factors as in (\ref{eq01}).
 Applying {\em determinant break-up} \cite{BREAKUP-H,BREAKUP-AH} one
 writes
\be\label{eq02}
\left( \det Q^2 \right)^\alpha = 
\left[\left( \det Q^2 \right)^{\alpha/n_B} \right]^{n_B}
 \ . 
\ee
 with some positive integer $n_B$.
 In what follows we always consider a single determinant factor with
 an effective power $\alpha$:
\be\label{eq03}
\left( \det Q^2 \right)^\alpha  \ , \hspace{2em}
\alpha = \frac{N_f}{2n_B} \ .
\ee
 If there are several such factors in the path integral then each of
 them can be separately taken into account in the same way.

 The basic ingredient of TSMB is a polynomial approximation
\be\label{eq04}
P(x) \simeq x^{-\alpha} \ , \hspace{3em} x \in [\epsilon,\lambda]
\ee
 where the interval $[\epsilon,\lambda]$ covers the eigenvalue spectrum
 of $Q^2$ on gauge configurations having a non-negligible weight in
 the path integral.
 The determinant factor in the Boltzmann weight can then be taken into
 account with L\"uscher's multi-boson representation.
 Assuming that the roots of the polynomial $P(x)$ occur in complex
 conjugate pairs, one can introduce the equivalent forms
\be\label{eq05}
P(Q^2)
= r_0 \prod_{j=1}^n [(Q \pm \mu_j)^2 + \nu_j^2]
= r_0 \prod_{j=1}^n (Q-\rho_j^*) (Q-\rho_j) \ ,
\ee
 where $n$ is the degree of $P(x)$ and the roots are
 $r_j \equiv (\mu_j+i\nu_j)^2$ with $\rho_j \equiv \mu_j + i\nu_j$.
 With the help of complex boson (pseudofermion) fields $\Phi_{jx}$
 one can write
\begin{eqnarray}\nonumber
\left( \det Q^2 \right)^\alpha 
& \propto &
\prod_{j=1}^n\det[(Q-\rho_j^*) (Q-\rho_j)]^{-1}
\\ \label{eq06}
& \propto &
\int [d\Phi]\; \exp\left\{ -\sum_{j=1}^n \sum_{xy}
\Phi_{jy}^+\, [(Q-\rho_j^*) (Q-\rho_j)]_{yx}\,
\Phi_{jx} \right\} \ .
\end{eqnarray}

 In the representation (\ref{eq06}) the complex boson fields
 $\Phi_{jx},\; j=1,2,\ldots,n$ carry the indices of the corresponding
 fermion fields.
 For instance, in QCD with Wilson-type fermions there are colour and
 Dirac-spinor indices.
 Since the multi-boson action in (\ref{eq06}) is local, similarly to
 the gauge field action, one can apply the usual bosonic updating
 algorithms like Metropolis, heatbath or overrelaxation.
 In fact, the multi-boson action is Gaussian hence for the multi-boson
 fields a global heatbath update is also possible which creates, for
 a fixed gauge field, a statistically independent new set of boson
 fields.

 The polynomial approximation in (\ref{eq04}) is not exact.
 In order to obtain an exact updating algorithm one has to correct for
 its deviation from the function to be approximated.
 One can easily show that for small fermion masses in lattice units the
 (typical) smallest eigenvalue of $Q^2$ behaves as $(am)^2$ and for a
 fixed quality of approximation within the interval 
 $[\epsilon,\lambda]$ the degree of the polynomial is growing as
 $n \propto \epsilon^{-1/2} \propto (am)^{-1}$.
 In general, the polynomial approximation has to be precise enough in
 order that the deviations in expectation values be smaller than the
 statistical errors.
 In practical applications, for instance in QCD simulations, this
 would require very high degree polynomials with $n$ of the order
 $10^3$-$10^4$.
 (For numerical examples showing the convergence rate of the polynomial
 approximations see \cite{POLYNOM2}.)
 Performing numerical simulations with such a high $n$ is practically
 impossible (and would be in any case completely ineffective).
 The way out is to perform the corrections stochastically.

 For improving the approximation in (\ref{eq04}) a second polynomial
 is introduced:
\be\label{eq07}
P_1(x) P_2(x) \simeq x^{-\alpha} \ , \hspace{2em} 
x \in [\epsilon,\lambda] \ .
\ee
 The first polynomial $P_1(x)$ gives a crude approximation as in
 (\ref{eq04}) with $P_1(x) \equiv P(x)$.
 The second polynomial $P_2(x)$ gives a good approximation according to
\be\label{eq08}
P_2(x) \simeq [x^\alpha P_1(x)]^{-1} \ .
\ee

 During the updating process $P_1$ is realized by multi-boson updates
 whereas $P_2$ is taken into account stochastically by a {\em noisy
 correction step}.
 For this, after preparing a new set of gauge fields $[U^\prime]$
 from the old one $[U]$ by local updates, one generates a Gaussian
 random vector having a distribution
\be \label{eq09}
\frac{e^{-\eta^\dagger P_2(Q[U]^2)\eta}}
{\int [d\eta] e^{-\eta^\dagger P_2(Q[U]^2)\eta}}  \ ,
\ee
 and accepts the change of the gauge field $[U] \rar [U^\prime]$ with
 probability
\be \label{eq10}
\min\left\{ 1,A(\eta;[U^\prime] \lar [U]) \right\} \ ,
\ee
 where
\be \label{eq11}
A(\eta;[U^\prime] \lar [U]) =
\exp\left\{-\eta^\dagger P_2(Q[U^\prime]^2)\eta
           +\eta^\dagger P_2(Q[U]^2)\eta\right\}\ .
\ee
 One can show \cite{TSMB} that this update procedure satisfies the
 detailed balance condition and hence creates the correct distribution
 of the gauge fields.
 (See the proof for the more general case of MSMB given below in
 (\ref{eq20})-(\ref{eq23}).)

 The Gaussian noise vector $\eta$ can be obtained from $\eta^\prime$
 distributed according to the simple Gaussian distribution
\be \label{eq12}
\frac{e^{-\eta^{\prime\dagger}\eta^\prime}}
{\int [d\eta^\prime] e^{-\eta^{\prime\dagger}\eta^\prime}}
\ee
 by setting it equal to
\be \label{eq13}
\eta = P_2(Q[U]^2)^{-\half} \eta^\prime  \ .
\ee
 In order to obtain the inverse square root on the right hand side of
 (\ref{eq13}), one can proceed with a polynomial approximation
\be \label{eq14}
 \bar{P}_2(x) \simeq P_2(x)^{-\half} \ , \hspace{1em}
x \in [\bar{\epsilon},\lambda] \ .
\ee
 Note that here the interval $[\bar{\epsilon},\lambda]$ can be chosen
 differently, usually with $\bar{\epsilon} < \epsilon$, from the
 approximation interval $[\epsilon,\lambda]$ for $P_2$.

 The polynomial approximation in (\ref{eq07}) can only become exact
 in the limit when the degree $n_2$ of the second polynomial $P_2$ is
 infinite.
 Instead of investigating the dependence of expectation values on $n_2$
 by performing several simulations, it is also possible to fix some high
 value of $n_2$ for the simulation and perform another correction in the
 {\em measurement} of expectation values by still finer polynomials.
 This is done by {\em reweighting} the configurations.
 (A similar reweighting procedure is applied in the PHMC algorithm of
 Ref.~\cite{PHMC}.)
 This {\em measurement correction} is based on a further polynomial
 approximation $P^\prime$ with polynomial degree $n^\prime$ which
 satisfies
\be\label{eq15}
\lim_{n^\prime \to \infty} P_1(x)P_2(x)P^\prime(x) =
x^{-\alpha} \ , \hspace{3em}
x \in [\epsilon^\prime,\lambda] \ .
\ee
 The interval $[\epsilon^\prime,\lambda]$ can be chosen by convenience,
 for instance, such that $\epsilon^\prime=0,\lambda=\lambda_{max}$,
 where $\lambda_{max}$ is an absolute upper bound of the eigenvalues of
 $Q^2$.
 (In case of $\epsilon^\prime=0$ the approximation interval is strictly
 speaking $(\epsilon^\prime,\lambda]$.
 An absolute upper bound for the eigenvalues of $Q^2$ exists because the
 commonly used fermion matrices are bounded from above.)
 In practice, instead of $\epsilon^\prime=0$, it is more effective to
 take $\epsilon^\prime > 0$ and determine the eigenvalues below
 $\epsilon^\prime$ and the corresponding correction factors exactly.
 For the evaluation of $P^\prime$ one can use $n^\prime$-independent
 recursive relations \cite{POLYNOM1}, which can be stopped by observing
 the required precision of the result.
 After reweighting the expectation value of a quantity $A$ is given by
\be\label{eq16}
\langle A \rangle = \frac{
\langle A \exp{\{\eta^\dagger[1-P^\prime(Q^2)]\eta\}}
\rangle_{U,\eta}}
{\langle  \exp{\{\eta^\dagger[1-P^\prime(Q^2)]\eta\}}
\rangle_{U,\eta}} \ ,
\ee
 where $\eta$ is a simple Gaussian noise like $\eta^\prime$ in
 (\ref{eq12}).
 Here $\langle\ldots\rangle_{U,\eta}$ denotes an expectation value
 on the gauge field sequence, which is obtained in the two-step process
 described before, and on a sequence of independent $\eta$'s.
 The expectation value with respect to the $\eta$-sequence can be
 considered as a Monte Carlo updating process with the trivial action
 $S_\eta \equiv \eta^\dagger\eta$.
 The length of the $\eta$-sequence on a fixed gauge configuration can,
 in principle, be arbitrarily chosen.
 In practice it has to be optimised for obtaining the smallest possible
 errors with a given amount of computer time.

 The polynomial approximations in (\ref{eq04}), (\ref{eq08}),
 (\ref{eq14}) and (\ref{eq15}) can be obtained in a recursive scheme
 providing least-square optimisation \cite{POLYNOM1,POLYNOM2}.
 Numerical methods to determine the polynomial coefficients can be
 based either on arbitrary precision arithmetics \cite{GEB-MONT}
 or on discretisation of the approximation interval \cite{KATZ-TOTH}.
 The expansion in appropriately defined orthogonal polynomials is
 an important ingredient, both in determining the polynomial
 coefficients and in the application of the polynomials of the squared
 fermion matrix $Q^2$ on a vector.

 Least-square optimisation corresponds to minimising the $L_2$-norm of
 the deviation.
 An often used alternative is to minimise the $L_\infty$ norm which is
 equivalent to the minimisation of the maximal relative deviation.
 In general, the goal is to obtain the smallest possible deviation of
 the expectation values with the smallest possible polynomial degree.
 The experience with the least-square optimisation in TSMB has been
 rather satisfactory because it gives the best overall fit of the
 lattice action with a given polynomial degree.
 (For numerical examples comparing $L_2$- with $L_\infty$-optimisation
 see Ref.~\cite{POLYNOM2}.)
 The often stated advantage of minimising the upper limit of the
 relative deviation of the lattice action is relativised by the fact
 that the deviation of the expectation values from the correct ones is
 in general a complicated function of the deviation in the lattice
 action.
 
 The {\em multi-step multi-boson} (MSMB) updating algorithm is a
 straightforward generalisation of TSMB updating.
 Instead of the two-step approximation in (\ref{eq07}) we now
 consider a sequence of polynomial approximations of arbitrary length:
\be\label{eq17}
P_1(x) P_2(x) \ldots P_k(x) \simeq x^{-\alpha} \ , \hspace{2em} 
x \in [\epsilon_k,\lambda] \ .
\ee
 Here the subsequent polynomials define approximations with increasing
 precision according to
\be\label{eq18}
P_i(x) \simeq [x^\alpha P_1(x)\ldots P_{i-1}(x)]^{-1} \ , \hspace{2em}
(i = 2,3,\ldots,k) \ .
\ee
 The first polynomial $P_1$ is realized during updating by local updates
 as in TSMB.
 The higher approximations $P_2,\ldots,P_k$ are implemented by a
 sequence of nested noisy correction steps as in
 (\ref{eq09})-(\ref{eq11}).
 The necessary Gaussian distributions of noise vectors can be obtained
 by appropriate polynomials, similarly to (\ref{eq14}):
\be \label{eq19}
 \bar{P}_i(x) \simeq P_i(x)^{-\half} \ , \hspace{1em}
(i = 2,3,\ldots,k) \ , \hspace{2em}
x \in [\bar{\epsilon}_k,\lambda] \ .
\ee

 The proof of the {\em detailed balance} condition for MSMB goes
 essentially in the same way as for TSMB.
 The aim is to reproduce with the first $i$ correction steps the
 canonical distribution of the gauge field
\be \label{eq20}
w_{(i)}[U] = e^{-S_g[U]} \left\{ \det P_1[U]\, \det P_2[U] \ldots
\det P_i[U] \right\}^{-1} \ , \hspace{1em} (i = 1,2,\ldots,k) \ ,
\ee
 where the short notation
 $P_i[U] \equiv P_i(Q[U]^2)$ is used and $S_g[U]$ denotes the action
 for the gauge field.
 Let us assume that detailed balance holds for the first $(i-1)$
 steps, that is the transition probability
 $P_{(i-1)}([U^\prime] \lar [U])$ satisfies
\begin{eqnarray}\nonumber
& & P_{(i-1)}([U^\prime] \lar [U])
e^{-S_g[U]}\, \left\{ \det P_1[U]\ldots\det P_{i-1}[U] \right\}^{-1} =
\\ \label{eq21}
& & P_{(i-1)}([U] \lar [U^\prime])
e^{-S_g[U^\prime]}\, \left\{ \det P_1[U^\prime]\ldots
\det P_{i-1}[U^\prime] \right\}^{-1} \ .
\end{eqnarray}
 The transition probability of the $i$'th step is a product of
 $P_{(i-1)}([U^\prime] \lar [U])$ with the acceptance probability
 $P_{(i)a}([U^\prime] \lar [U])$:
\be \label{eq22}
P_{(i)}([U^\prime] \lar [U]) = P_{(i-1)}([U^\prime] \lar [U])\,
P_{(i)a}([U^\prime] \lar [U]) \ .
\ee
 It is easy to show that if $P_{(i)a}([U^\prime] \lar [U])$ is
 defined according to (\ref{eq09})-(\ref{eq11}) with $P_2$ replaced
 by $P_i$ then the acceptance probability satisfies
\be \label{eq23}
P_{(i)a}([U^\prime] \lar [U])\,\left\{\det P_i[U]\right\}^{-1} =
P_{(i)a}([U] \lar [U^\prime])\,\left\{\det P_i[U^\prime]\right\}^{-1}
\ .
\ee
 From this immediately follows that the transition probability
 of the $i$'th step \\
 $P_{(i)}([U^\prime] \lar [U])$ satisfies the detailed balance
 condition (\ref{eq21}) with $(i-1)$ replaced by $(i)$.

 An alternative way to prove that the described procedure creates the
 correct distribution of the gauge fields is to consider the fields
 $\eta$ as additional pseudofermion fields in the Markov chain with
 the lattice action given by the exponent in (\ref{eq09}).

 The advantage of the multi-step scheme compared to the two-step one
 is that the lower approximations can be chosen to be less accurate
 and consequently have lower polynomial degrees and are faster to
 perform.
 The last approximations, which are very precise and need high
 polynomial degrees, can be done less frequently.
 The last polynomial $P_k$ can already be chosen so precise that,
 for some given statistical error, the measurement correction with
 $P^\prime$ becomes unnecessary.

 An easy generalisation of the multi-step scheme described until now
 is to require the correct function to be approximated in (\ref{eq17})
 only in the last step and allow for functions easier to approximate
 in the previous steps.
 This means that (\ref{eq18}) can be generalised, for instance, to
\be\label{eq24}
P_i(x) \simeq [(x+\rho_i)^\alpha P_1(x)\ldots P_{i-1}(x)]^{-1} \ ,
\hspace{2em} (i = 1,2,\ldots,k) \ .
\ee
 with positive $\rho_i$ and $\rho_k=0$.
 This has a resemblance to the ``mass preconditioning'' as introduced
 for HMC algorithms in Ref.~\cite{HASENBUSCH,HASENBUSCH-JANSEN}.
 The advantage of (\ref{eq24}) is that for $\rho_i > 0$ one can decrease
 the degree of the polynomial $P_i(x)$ and at the same time, if
 $\rho_i/\rho_{i-1}$ is not much smaller than 1, the acceptance in
 the $i$'th correction step remains high enough.

 There are other multi-step approximation schemes conveivable: for
 instance, one can take $P_i(x) \simeq x^{-\alpha/k},\; (i=1,\ldots,k)$
 which corresponds to the determinant breakup in (\ref{eq02}).
 Similarly, ``mass preconditioning'' can also be considered as a 
 generalisation of determinant breakup.

 We performed several tests with the MSMB algorithms in some of the
 simulation points of Ref.~\cite{DBW} with the Wilson fermion action
 for two flavours of quarks and the DBW2 gauge action \cite{DBW2} for
 the colour gauge field.
 In particular, on an $8^3 \cdot 16$ lattice at
 $\beta=0.55,\;\kappa=0.188,\;\mu=0$ (simulation point $(c)$ in
 \cite{DBW} with a bare quark mass in lattice units $am_q \simeq 0.015$)
 a three-step algorithm was tuned for obtaining better performance.
 (Here $\mu$ denotes the ``twisted mass'' which is actually set equal
 to zero in these runs.)
 In another test run on a $16^3 \cdot 32$ lattice we have chosen a point
 where a detailed simulation has been performed recently with both the
 TSMB and HMC algorithm \cite{TOBEPUB}, namely at
 $\beta=0.74,\;\kappa=0.158,\;\mu=0$ with a bare quark mass in lattice
 units $am_q \simeq 0.024$.
 In a three-step algorithm the following parameters were chosen:
 $n_B=2,\; n_1=60,\; n_2=200,\; \bar{n}_2=300,\; n_3=800,\;
 \bar{n}_3=900$.
 (The degree of the polynomials $P_i$ and $\bar{P}_i$ is denoted by
 $n_i$ and $\bar{n}_i$, respectively.) 
 The second correction step was called after performing 10 update
 cycles involving the first correction.
 The integrated autocorrelation for the average plaquette in these test
 runs were typically around $\tau_{\rm plaq}^{\rm int} \simeq 10$
 full update cycles including the second correction.

 The simulation costs in these runs turned out to be, even with a
 moderate effort put in parameter tuning, by about a factor of 1.5 lower
 than in the corresponding well-tuned TSMB runs.
 The gain comes from the lower cost of the first correction compared
 to the correction step in TSMB.
 The cost of the second correction does not contribute much to the
 full cost because it is done infrequently.
 For instance, on the $16^3 \cdot 32$ lattice the TSMB run had the
 parameters $n_B=4,\;n_1=34,\;n_2=720,\;\bar{n}_2=740$.
 (Note that the cost of the correction is mainly determined by the
 product $n_B(n_2+\bar{n}_2)$ which is 5840 in TSMB and only
 1000 in the first correction of MSMB.)

%%%%%%%%%%%%%%%%%%%%%%%%%%%%%%%%%%%%%%%%%%%%%%%%%%%%%%%%%%%%%%%%%%%%%%%%
\section{Multi-step correction for HMC}\label{sec3}

 The first (updating) step producing a new gauge field configuration
 can also be replaced by Hybrid Monte Carlo trajectories \cite{HMC}.
 In this step some approximation of the fermion determinant can be
 used and after a few trajectories one can perform a stochastic
 correction step.
 The rest within a multi-step correction scheme is the same as in MSMB
 updating.

 A possible application of multi-step stochastic corrections is to
 perform a HMC update with a mass-preconditioned fermion matrix which
 corresponds to $\rho_1 > 0$ in Eq.~(\ref{eq24}) and correct for the
 exact determinant (that is, $\rho_1 = 0$)  stochastically.
 The polynomials for the stochastic corrections are defined in the same
 way as in (\ref{eq24}).

 Another possibility is to start by an update step as in polynomial
 hybrid Monte Carlo (PHMC) \cite{PHMC}.
 In order to generate the correct distribution of pseudofermion fields
 at the beginning of the trajectory one needs a polynomial as in
 (\ref{eq19}) also for $i=1$:
\be \label{eq25}
 \bar{P}_1(x) \simeq P_1(x)^{-\half} \ , \hspace{3em}
x \in [\epsilon,\lambda] \ .
\ee
 In order to avoid very high degree first polynomials $P_1(x)$,
 which would cause problems with rounding errors in the calculation
 of the fermionic force \cite{ORDER}, one should use determinant
 break-up (see Eq.~(\ref{eq02})).
 The ordering of the root factors in the expression of the fermionic
 force \cite{PHMC} is best done according to the procedure proposed in
 \cite{POLYNOM1}.
 Again, the stochastic correction steps can be performed during the update
 according to the procedure described in Section \ref{sec2}.

 Besides decreasing the polynomial degrees in the PHMC update step,
 another advantage of applying determinant breakup is that both
 magnitude and variance of the quark force is decreased approximately
 proportional to $n_B^{-1/2}$.

 In some test runs on $8^3 \cdot 16$ lattices the performance of the
 PHMC algorithm with stochastic correction turned out to be promisingly
 good.
 In particular, we performed simulations with the parameters
 $\beta=0.55,\;\kappa=0.184,\;0.186,\;0.188,\;\mu=0$
 corresponding to the points $(a)$, $(b)$ and $(c)$ in Ref.~\cite{DBW}
 with bare quark masses in lattice units
 $am_q \simeq 0.071,\; 0.039,\; 0.015$, respectively.
 The PHMC trajectories were created by applying the
 Sexton-Weingarten-Peardon integration scheme with multiple time scales
 \cite{SEXWEIN,PEARSEX}.
 Gains up to factors of 5 were observed in comparison with the
 costs of the TSMB runs.
 The origin of this better performance is that the integrated
 autocorrelations are shorter, whereas the costs for one update cycle
 are similar to TSMB (see Table 3 of \cite{DBW}).
 These numbers also show that in these points PHMC with stochastic
 correction is better than MSMB.

%%%%%%%%%%%%%%%%%%%%%%%%%%%%%%%%%%%%%%%%%%%%%%%%%%%%%%%%%%%%%%%%%%%%%%%%
\section{Summary}\label{sec4}

 In summary, multi-step stochastic correction is a useful and flexible
 tool which can be implemented in both multi-bosonic and hybrid Monte
 Carlo update algorithms.
 In the present paper we reported on first tests with the multi-step
 multi-boson (MSMB) and stochastically corrected polynomial Monte
 Carlo algorithm which look promising.
 In our test runs on relatively small lattices and with moderately small
 quark masses the PHMC algorithm with stochastic correction is faster
 than MSMB.
 Of course, further tests on larger lattices and at smaller quark masses
 are necessary before applying these updating algorithms in large scale
 simulations.
 The relation between the cost factors of MSMB versus PHMC may also be
 different depending on the lattice volume and quark mass.

 Based on our experience with the TSMB algorithm, we expect the
 computational costs of our multi-step stochastic correction schemes to
 increase only slightly faster than linear with the number of lattice
 sites.
 This differs from the multi-level Metropolis scheme proposed
 in Ref.~\cite{HASENBUSCH,MULTIMET} where the volume dependence is
 quadratic.

 An important feature of both the MSMB and of the PHMC algorithm with
 multi-step stochastic correction is that they are applicable for odd
 numbers of flavours, too, provided that there is no sign problem with
 the fermion determinant.
 The same holds for the rational hybrid Monte Carlo (RHMC) algorithm
 \cite{RHMC} where multi-step stochastic correction might also be
 useful.

 The main advantage of the stochastic correction in several steps
 compared to a single stochastic correction is that the costly last
 correction has to be done infrequently.
 This feature becomes increasingly more important for large lattices
 at small fermion masses where the cost of the last correction
 increases proportional to the inverse quark mass in lattice units.

\newpage%\vspace*{1.5em}
\noindent
{\large\bf Acknowledgements}

\noindent
 We thank the authors of the papers of Ref.~\cite{TWIST,DBW}, in
 particular Roberto Frezzotti, Karl Jansen and Carsten Urbach, for
 enlightening discussions on fermion updating algorithms.
 The computations were performed on the IBM-JUMP computer at NIC
 J\"ulich and the PC clusters at DESY-Hamburg.

%%%%%%%%%%%%%%%%%%%%%%%%%%%%%%%%%%%%%%%%%%%%%%%%%%%%%%%%%%%%%%%%%%%%%%%%

\vspace*{2em} %\newpage
%%%%%%%%%%%%%%%%%%%%%%%%%%%%%%%%%%%%%%%%%%%%%%%%%%%%%%%%%%%%%%%%%%%%%%%%

%%%%%%%%%%%%%%%%%%%%%%%%%%%%%%%%%%%%%%%%%%%%%%%%%%%%%%%%%%%%%%%%%%%%%%%%

\end{document}